%% file: main.tex
\documentclass[10pt,conference]{IEEEtran}
\IEEEoverridecommandlockouts
\usepackage{amsmath,amssymb,amsfonts}
\usepackage{comment}
\usepackage{cite}
\usepackage{graphicx}
\usepackage{textcomp}
\usepackage[breaklinks,colorlinks,linkcolor=blue,citecolor=blue,urlcolor=blue]{hyperref}
\usepackage{colortbl}
\usepackage{multirow}
\usepackage{footnote}
\usepackage{tablefootnote}
\usepackage{float}
\usepackage{threeparttable}
\usepackage{pifont}
\usepackage{textcomp}
\usepackage{tabularx}
\usepackage[table]{xcolor}
\usepackage{placeins}
\usepackage{booktabs}

\usepackage{algorithmic}
\usepackage[ruled,vlined,linesnumbered,commentsnumbered]{algorithm2e}
\usepackage{arydshln}

\def\BibTeX{{\rm B\kern-.05em{\sc i\kern-.025em b}\kern-.08em
    T\kern-.1667em\lower.7ex\hbox{E}\kern-.125emX}}

\begin{document}

\newcommand{\ml}[1]{{\color{red}\bf [Meng: #1]}}
\newcommand{\yx}[1]{{\color{blue}\bf [Yixuan: #1]}}
\newcommand{\zjw}[1]{{\color{green}\bf [zjw: #1]}}
\newcommand{\xt}[1]{{\color{black} #1}}
\newcommand{\xietong}[1]{{\color{purple}\bf [xt: #1]}}
\newcommand{\zw}[1]{{\color{magenta}[Zishen: #1]}}

\newcommand{\red}[1]{{\color{red}\bf (#1)}}
\newcommand{\magenta}[1]{{\color{magenta} #1}}
\newcommand{\method}{ReaLM}

\title{Aging Aware Adaptive Voltage Scaling for Reliable and Efficient AI Accelerators
\vspace{-15pt}
}

\author{
Tong Xie$^{21}$,
Zuodong Zhang$^{4}$,
Chao Yang$^{5}$, 
Yuan Wang$^{23}$,
Runsheng Wang$^{234}$,
and Meng Li$^{123 \dag }$
\\
\textit{$^1$Institute for Artificial Intelligence \& $^2$School of Integrated Circuits, Peking University, Beijing, China} \\
\textit{$^3$Beijing Advanced Innovation Center for Integrated Circuits, Beijing, China} \\
\textit{$^4$Institute of Electronic Design Automation, Peking University, Wuxi, China} \\
\textit{$^{5}$T-Head Semiconductor Co., Ltd, Shanghai, China}

\vspace{-20pt}
\thanks{

$^\dag$Corresponding author: meng.li@pku.edu.cn. 

This work was supported in part by NSFC under Grant 62495102,  Grant 92464104, and Grant 62125401, in part by Beijing Outstanding Young Scientist Program under Grant JWZQ20240101004, in part by Beijing Municipal Science and Technology Program under Grant Z241100004224015, and in part by 111 Project under Grant B18001.}
}

\vspace{-20pt}


\newcommand{\opt}{\texttt{OPT-1.3B}}
\newcommand{\llama}{\texttt{LLaMA-2-7B}}
\newcommand{\llm}{\texttt{LLaMA-3-8B}}
\newcommand{\mistral}{\texttt{Mistral-7B}}

\maketitle
\input{docs/0_Abstract}
\input{docs/1_Introduction}
\input{docs/2_Background}

\input{docs/3_Framework}
\input{docs/4_Optimization}
\input{docs/6_Evaluation}

\input{docs/7_Conclusion}


\bibliographystyle{ieeetr}

\bibliography{top_simplified.bib,reference_simplified_modified.bib}
\end{document}

%% file: docs/0_Abstract.tex
\begin{abstract}
Deep neural networks (DNNs) have showcased remarkable performance across various tasks and are widely deployed on AI accelerators fabricated in advanced technology nodes for efficiency. As aging effects become more pronounced, timing and voltage guardbands are increasingly applied. Aging-aware adaptive voltage scaling (AVS), which adjusts supply voltage based on on-chip aging scenarios, has emerged as a promising solution to avoid excessive guardbanding. However, conventional AVS techniques overlook the inherent resilience of DNNs and frequently raise the supply voltage unnecessarily, thereby exacerbating aging and increasing power consumption. 
To enable reliable and efficient AI inference with AVS, in this paper, we develop an accurate aging prediction framework that incorporates historical effects and iterative extrapolation for full-lifetime modeling. 
Building on this framework, we propose a fault-tolerant voltage scaling policy that exploits DNN resilience and defers voltage increases accordingly. Experiments show that, our framework mitigates the pessimism of maximum-voltage baselines, reducing predicted threshold voltage shift ($\Delta V_{\rm th}$) by 19.4\% for PMOS and 19.1\% for NMOS, respectively. Furthermore, evaluation on representative DNN workloads demonstrates that our optimization reduces aging degradation by up to 45.8\% (NMOS) and 30.6\% (PMOS) while achieving 14.0\% average lifetime power savings compared to resilience-agnostic methods.

\end{abstract}


\begin{IEEEkeywords}
Adaptive voltage scaling; aging; fault tolerance
\end{IEEEkeywords}

%% file: docs/1_Introduction.tex
\section{Introduction}
\label{sec: intro}

Recent advances in deep neural networks (DNNs), particularly large language models (LLMs), have revolutionized a wide range of applications \cite{zhao2023survey}
To meet their enormous computational demands‌, these models are increasingly deployed on customized AI accelerators such as TPU-like systolic arrays \cite{jouppi2017datacenter}, which are typically fabricated in advanced technology nodes for efficient inference. However, as Moore’s Law scales toward the nanoscale, aging effects, such as bias temperature instability (BTI) \cite{guo2017towards, schroder2007negative, grasser2009two, ghosh2010parameter} and hot carrier injection (HCI) \cite{yu2017new, takeda2005empirical, sun2023transient}, become increasingly pronounced. These effects manifest as increased threshold voltage ($\Delta V_{\rm th}$) and potential timing errors, posing challenges to ensuring computational correctness \cite{dixit2021silent, moghaddasi2023dependable, jiao2017clim, huang2017variability}. Therefore, how to enable reliable yet efficient AI acceleration becomes an important question and attracts increasing attention \cite{xie2025realm}. 

To accommodate aging effects and guarantee functional correctness over the product lifetime, static timing analysis (STA) is employed to introduce timing or voltage guardbands. However, these guardbands are often pessimistically estimated for worst-case scenarios, resulting in significant overdesign costs \cite{zhang2022avatar}. Such costs escalate with technology scaling and may ultimately diminish gains from device scaling \cite{rahimi2016variability, zhang2023read}.

\begin{figure}[!tb]
    \centering
    \includegraphics[width=1\linewidth]{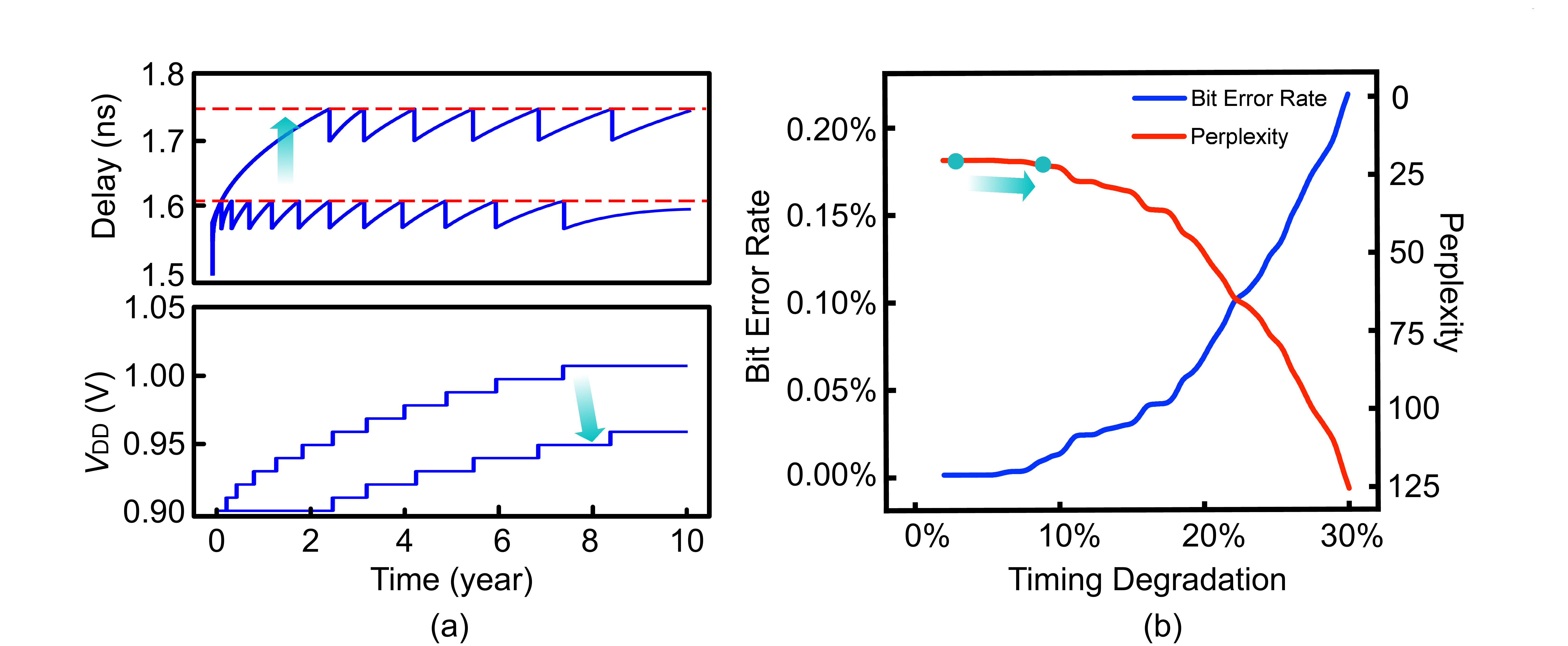}
    \vspace{-20pt}
    \caption{(a) AVS compensates for aging-induced path delay by increasing the supply voltage. (b) Aging-induced timing degradation raises BER, impairing DNN performance. However, the inherent error resilience of DNNs allows certain timing violations, which can, in turn, improve AVS efficiency. Perplexity is measured using \opt~on the WikiText-2 dataset.
    }
    \label{fig:1}
    \vspace{-15pt}
\end{figure}

Adaptive voltage scaling (AVS) is a promising technique for mitigating aging guardbands by adaptively increasing the supply voltage ($V_{\mathrm{DD}}$) in response to aging-induced circuit degradation \cite{mintarno2011self, cho2015aging, huard2016aging, cho2016postsilicon, chan2014aging}. As illustrated in Fig. \ref{fig:1}(a), AVS leverages on-chip sensors to monitor aging and timing, enabling dynamic voltage adjustment and eliminating the need for pessimistic worst-case guardbands. 

Despite its efficacy in reducing aging guardbands, AVS still faces efficiency challenges. First, existing AVS designs estimate aging based on the maximum supply voltage \cite{cho2016postsilicon, chan2014aging}, leading to significant overestimation and suboptimal power, performance, and area (PPA) overhead. Second, classical AVS incrementally increases $V_{\mathrm{DD}}$ multiple times during a product’s lifetime, which exacerbates power consumption and accelerates aging. For example, over a 10-year lifetime, $V_{\mathrm{DD}}$ may rise from 0.90 V to 1.02 V, resulting in over 100 mV of threshold voltage shift. This is because classical AVS increases $V_{\rm DD}$ for every detected timing violation, regardless of its actual impact. As shown in Fig. \ref{fig:1}(b), DNN performance remains stable under lower bit error rates (BERs), revealing inherent error resilience. Exploiting this resilience by adaptively allowing timing violations in the AVS process can reduce unnecessary voltage increases, thus enhancing both reliability and energy efficiency, as demonstrated in Fig. \ref{fig:1}(a).

To address these challenges, we develop an accurate aging evaluation framework for AVS systems that models BTI and HCI effects under voltage variations, incorporating historical voltage dependency and iterative extrapolation for 10-year lifetime prediction (Sec. \ref{sec:avs}). 
Building on this framework, we optimize the AVS voltage adjustment policy by exploiting the inherent error resilience of DNNs. Specifically, we establish quantitative relationships between aging and BER, and between BER and model accuracy, enabling controlled deferral of voltage adjustments  while meeting user-specified accuracy targets (Sec. \ref{sec: optimization}). Our contributions are summarized as follows.


\begin{itemize}
    \item We develop an accurate aging evaluation framework for AVS systems by incorporating advanced BTI and HCI models. Unlike traditional approaches limited to constant voltage stress, our framework supports aging prediction under arbitrary voltage waveforms, making it well-suited for AVS scenarios. We further employ iterative extrapolation to enable full-lifetime aging prediction.
    \item Building on our evaluation framework, we propose fault-tolerant voltage scaling, a voltage-adjustment policy optimized for AI accelerators running DNN workloads. By exploiting the inherent fault tolerance of DNNs and permitting certain timing violations, this approach mitigates lifetime voltage increases over lifetime while perserving DNN accuracy.
    \item Experimental results show that our framework enables accurate evaluation of aging-induced threshold-voltage shift ($\Delta V_{\rm th}$) throughout the AVS process. Compared with a maximum-voltage-based aging estimate, it reduces pessimism, yielding 19.4\% and 19.1\% lower estimated $\Delta V_{\rm th}$ for PMOS and NMOS, respectively. Using \llm~as a case study, we further show that our proposed policy reduces the aging degradation by up to 45.8\% (NMOS) and 30.6\% (PMOS), while delivering a 14.0\% average lifetime power reduction on representative workloads.

\end{itemize}

%% file: docs/2_Background.tex
\section{Background}

\subsection{Aging Mechanisms}
\begin{figure}[!tb]
    \centering
    \includegraphics[width=1\linewidth]{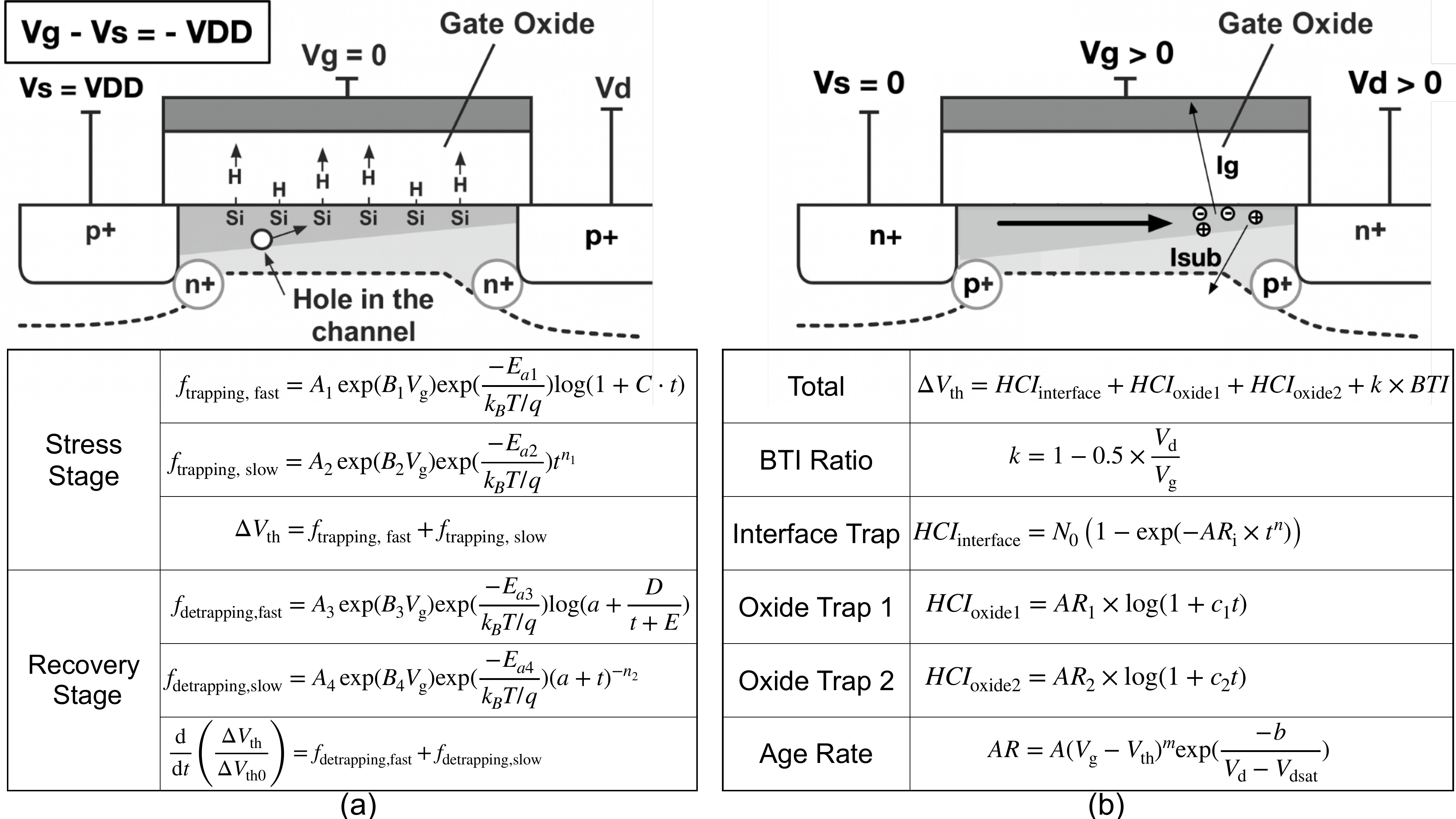}
    \vspace{-20pt}
    \caption{Mechanisms and compact models of (a) BTI and (b) HCI, where $V$, $T$, and $t$ denote voltage, temperature, and time, respectively. $B_{1-4}$ are voltage acceleration factors, $E_{a1-4}$ are activation energies, $k_B$ is the Boltzmann constant, and $q$ is the elementary charge. All other parameters are technology-dependent.
    }
    \label{fig:aging}
    \vspace{-10pt}
\end{figure}

\xt{
Transistor aging has become increasingly significant with CMOS technology scaling to the 16/14 nm node and beyond. 
Aging effects can manifest as an increase in $V_{\rm th}$ and degrade transistor driving capability, leading to increased circuit delay and elevated timing error rates over time \cite{huang2017variability, zhang2022avatar, huard2016aging, zhang2023read, chan2014aging}.
The two primary aging mechanisms are BTI \cite{guo2017towards, schroder2007negative, grasser2009two, ghosh2010parameter} and HCI \cite{yu2017new, takeda2005empirical, sun2023transient}.
In digital circuits like accelerators, BTI is induced during periods of logic stability, whereas HCI occurs during dynamic toggling between 0 and 1.

BTI arises when a voltage stress is applied to the MOSFET gate, causing charge trapping in the gate oxide and resulting in a shift in $V_{\mathrm{th}}$. This degradation is partially recoverable when the stress is removed. Recent studies \cite{guo2017towards, sun2023transient} attribute BTI behavior to fast traps and slow traps, as depicted in Fig. \ref{fig:aging}(a).


HCI occurs during voltage transitions, when high electric fields accelerate channel carriers to gain sufficient kinetic energy to overcome the gate dielectric potential barrier. This injection of those \textit{hot} carriers into the gate dielectric causes $V_{\rm th}$ degradation as well. Fig. \ref{fig:aging}(b) exhibits unified compact models considering the interface trap and oxide traps in HCI proposed in \cite{yu2017new}.
}

\subsection{Adaptive Voltage Scaling}

AVS dynamically adjusts $V_{\rm DD}$ according to on-chip sensors, thus mitigating excessive worst-case guardbands. It initializes $V_{\rm DD}$ to a low voltage $V_{\rm init}$ and increments it by $V_{\rm step}$ upon sensor-detected timing violations, compensating for aging-induced delay degradation.  
While AVS reduces overdesign by adapting to variation and aging, thus eliminating static guardbands, two efficiency challenges remain: (1) \textbf{Aging overestimation}: AVS continuously adjusts voltage, but accurate aging modeling under advanced nodes must account for historical effects of aging. Current AVS implementations \cite{mintarno2011self, cho2015aging, huard2016aging, chan2014aging} typically estimate aging based on the maximum voltage, which often overestimates device aging. (2) \textbf{Frequent but unnecessary voltage increases}. Existing AVS increases $V_{\rm DD}$ upon each detected timing violation, which may not always be necessary, as neural networks typically exhibit inherent resilience. While voltage compensation mitigates aging temporarily, it accelerates further aging, creating a vicious cycle that increases power consumption and reduces circuit lifetime. Consequently, optimizing AVS strategy with accurate aging evaluation can lead to significant PPA improvements.

\subsection{Fault Tolerance of DNNs}
\label{sec: fault tolerance of DNN}


Unlike general-purpose computing, where a single bit flip can trigger catastrophic failures, DNNs exhibit inherent fault tolerance. In these models, individual bit errors typically manifest as minor numerical deviations rather than critical logic faults. Extensive research has quantified this resilience through error injection experiments, establishing the relationship between bit error rate (BER) and model performance. For instance, \cite{reagen2018ares, mahmoud2021optimizing} demonstrated that convolutional neural networks can sustain quasi-error-free performance under a BER of approximately $1\times 10^{-4}$. Similarly, \cite{xie2025realm} characterized the heterogeneous resilience of LLMs, observing that tolerable BER thresholds vary from $1\times 10^{-7}$ to $1\times 10^{-3}$ across different operators. These findings underscore the potential to strategically trade off minor computational errors for significant efficiency gains.

%% file: docs/3_Framework.tex
\section{Aging Evaluation Framework of AVS}
\label{sec:avs}

\begin{figure}[!tb]
    \centering
    \includegraphics[width=1\linewidth]{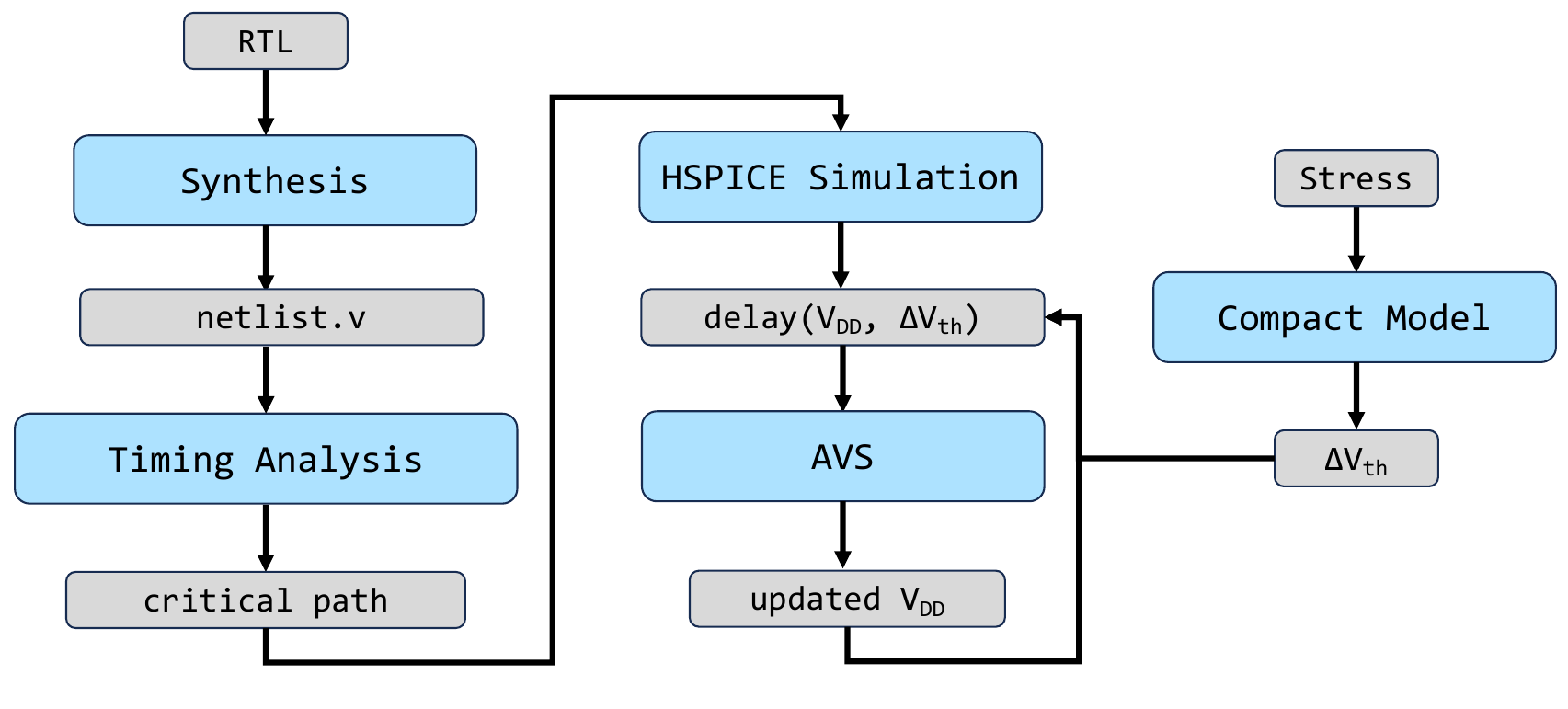}
    \vspace{-15pt}
    \caption{The proposed aging evaluation framework of AVS.
    }
    \label{fig:aging_evaluation_workflow}
    \vspace{-10pt}
  
\end{figure}

In this section, we present our aging evaluation framework of the AVS process. While prior approaches often overestimate aging by assuming maximum voltage, our framework leverages advanced and calibrated compact aging models~\cite{guo2017towards, yu2017new} of BTI and HCI, with coupling self heating effect also considered \cite{sun2023transient}. We further account for historical voltage adjustments and employ iterative extrapolation to enable efficient aging prediction over a 10-year lifetime. The workflow is depicted in Fig. \ref{fig:aging_evaluation_workflow}, including the following steps.
\begin{itemize}
    \item \textbf{Logic Synthesis}: Given the register transfer level (RTL) description of the target circuit, we use Synopsys Design Compiler to generate the gate-level netlist.
    \item  \textbf{Timing Analysis}: The synthesized netlist is analyzed using PrimeTime to extract critical paths, which are then converted into \texttt{.sp} files for HSPICE simulation.
    \item \textbf{HSPICE Simulation}: We simulate the extracted critical path delays and transition times under varying $\Delta V_{\rm th}$ and $V_{\rm DD}$ to obtain accurate delay characteristics.
    \item \textbf{Delay Modeling}: Polynomial fitting is applied to model the relationship between path delay, $\Delta V_{\rm th}$, and $V_{\rm DD}$.
    \item \textbf{$\Delta V_{\rm th}$ Estimation}: Aging-induced $\Delta V_{\rm th}$ are estimated using compact aging models \cite{yu2017new, guo2017towards}, based on the voltage stress profile from the AVS process. Note that the specific aging model can be readily replaced to match a given technology node without affecting the overall workflow of our framework.
    \item \textbf{AVS process}: During the simulated lifetime (e.g., 10 years), $V_{\rm DD}$ is iteratively increased by $V_{\rm step}$ whenever delay exceeds timing constraints, capturing the adaptive voltage scaling behavior over time.
\end{itemize}

\subsection{Logic Synthesis}

In this step, we take the RTL file of the target circuit as input. Specifically, we focus on a systolic array with 8-bit multipliers and a 32-bit accumulator in this paper. The technology library and clock period should be specified, whereas we adopt a commercial 14 nm FinFET PDK with a nominal voltage of 0.9 V and ambient temperature of 25 $\rm ^\circ $C, aligned with the aging model, and set the clock cycle to 1.6 ns. These inputs are sent to Synopsys Design Compiler via a synthesis script, which generates the gate-level netlist.


\begin{figure*}[t] 
    \centering
    \includegraphics[width=0.9\linewidth]{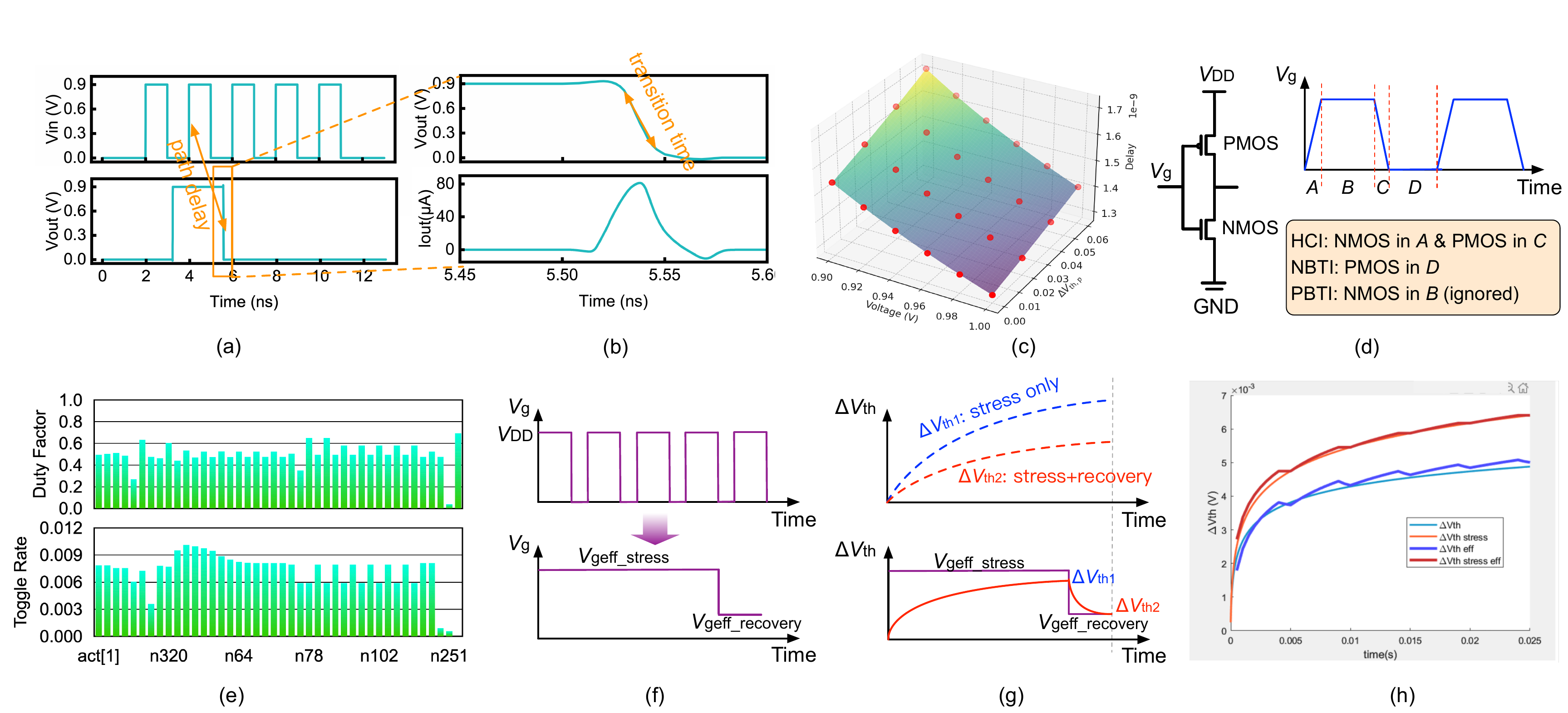}
    \vspace{-15pt}
    \caption{
    (a)(b) Simulated waveforms in HSPICE. (c) Fitted polynomial for path delay as a function $V_{\rm th, p}$ and $V_{\rm DD}$ with $V_{\rm th, n}=0$. (d) CMOS BTI and HCI mapping. (e) Duty factor and toggle rate along a representative critical path. (f)(g) Equivalent waveform with $N\times$ longer period, which satisfies $\Delta V_{\rm th1} = f_{\rm trapping}(V_{\rm geff\_stress}, t\times \mathrm {duty\_factor}) $ and $\Delta V_{\rm th2} = f_{\rm detrapping}(\Delta V_{\rm th1}, V_{\rm geff\_recovery}, t\times (1-\mathrm {duty\_factor)})$. (h) Iterative extrapolation of aging simulation.
    } 
    \label{fig:avs_figures}
    \vspace{-10pt}
\end{figure*}

\subsection{Timing Analysis}

The synthesized netlist is analyzed using PrimeTime to perform detailed timing analysis, enabling identification of the critical path. To ensure accuracy, parasitic capacitance is taken into account during the analysis. By specifying the subcircuit file, the critical path can be extracted using the \texttt{write\_spice\_deck} command, which generates both the SPICE-format circuit description and the corresponding stimulus file.
To reduce randomness, we export the 100 worst-case timing paths for further analysis.

\subsection{HSPICE Simulation}

To support AVS and aging analysis, we use HSPICE to accurately characterize path delay and transition time under various $\Delta V_{\rm th}$ and $V_{\rm DD}$. As illustrated in Figs. \ref{fig:avs_figures}(a)(b), path delay is measured from the clock signal reaching 50\% of $V_{\rm DD}$ to the corresponding output falling through 50\% of $V_{\rm DD}$, while transition time is defined as the interval between 10\% and 90\% of $V_{\rm DD}$. The library, parasitic capacitance, and subcircuit files are consistent with those used in the previous step. We parameterize the threshold voltage shifts of PMOS and NMOS ($\Delta V_{\rm th,p}$, $\Delta V_{\rm th,n}$) in the parasitic capacitance file and input voltage $V_{\rm DD}$ in the HSPICE stimulus file. These parameters are swept over defined ranges with fixed step sizes to get the path delay under various scenarios.
The reported delay is the average across the 100 worst paths.

\subsection{Delay Modeling}
\label{sec: polynomial}
Based on the simulated path delays across varying $V_{\rm th,n}$, $V_{\rm th,p}$, and $V_{\rm DD}$, we fit a ternary sixth-degree polynomial with 280 coefficients as a proximal model of $delay(V_{\rm th, p}, V_{\rm th, n}, V_{DD})$. 
As shown in Fig.~\ref{fig:avs_figures}(c), the fitted delay surface (with $V_{\rm th,n} = 0$) closely matches the HSPICE results (marked as red dots). The root mean square error (RMSE) is $5.85 \times 10^{-5}$ ns, which is orders of magnitude smaller than the nominal path delay ($\sim$1.5 ns), validating the accuracy of the fitted model.

\subsection{$\Delta V_{\rm th }$ Estimation}
\label{sec: delta_Vth}

We consider NBTI and HCI effects for PMOS and HCI for NMOS using the compact models in Fig. \ref{fig:aging}. PBTI for NMOS is ignored due to its insignificance \cite{ghosh2010parameter}. Fig. \ref{fig:avs_figures}(d) illustrates an example waveform and aging occurrence. 
To accurately characterize voltage stress under realistic workloads, we extracted input traces from a neural network inference pipeline and analyzed the duty factor and toggle rate of critical paths using Synopsys VCS. We performed simulations over 512k cycles to ensure statistical significance \cite{zhang2022avatar}. As illustrated in Fig. \ref{fig:avs_figures}(e), the duty factor for the majority of cells along a representative critical path ranges from 0.4 to 0.6, with toggle rates between 0.006 and 0.009, suggesting comparable $\Delta V_{\rm th}$ degradation profiles. We also observe similar trends across other critical paths. Since path delay accumulates over multiple stages, the effective degradation tends to converge toward the mean. Consequently, we use the average duty factor and toggle rate in subsequent evaluations to avoid the prohibitive overhead of modeling non-uniform cell-level aging.


For BTI, simulating its degradation cycle-by-cycle over years is computationally prohibitive. To address this, we begin with short-cycle basic waveforms and construct an equivalent waveform with a period $N$ times longer. We model alternating stress and recovery phases, where the stress duration is $t_{\rm stress} = \dfrac{t_{\rm clk}}{\rm{toggle\_rate}} \times \rm{duty\_factor}$ with $V_{\rm g} = V_{\rm DD}$, and the recovery duration is $t_{\rm recovery} = \dfrac{t_{\rm clk}}{\rm{toggle\_rate}} \times (1 - \rm{duty\_factor})$ with $V_{\rm g} = 0$. This enables degradation estimation over multiple cycles. As shown in Fig.~\ref{fig:avs_figures}(e)(f)(g), we iteratively extrapolate an equivalent long-period waveform to approximate the aging effect of $N$ short cycles. The effective stress voltage $V_{\rm geff\_stress}$ yields the same BTI degradation when considering only stress, while the effective recovery voltage $V_{\rm geff\_recovery}$ is nonzero and chosen to match the recovery behavior of the original waveform. By repeating this process, we can efficiently simulate BTI degradation over extended time scales.

In HCI, the transition current varies continuously, as illustrated in Fig. \ref{fig:avs_figures}(b). Unlike BTI, HCI exhibits no recovery effect, so the HCI process can be divided into smaller intervals and accumulated directly. For each voltage transition, we partition the switching period into $n$ small intervals and derive an effective HCI degradation factor $\gamma$ derived satisfying: 
$$
HCI(\gamma \times \mathrm{transition\_time}, V_{\rm DD}) = \sum\limits_{i=1}^{n}HCI(t_i-t_{i-1}, V_{\rm g}(t_{i}))
$$
The total HCI degradation time across cycles is then accumulated as:
$$
t_{HCI} = \gamma \times\dfrac{\mathrm{transition\_time}}{t_{\rm clk}}\times \mathrm {toggle\_rate} \times \mathrm {total\_time}
$$

\subsection{AVS Process}

As time progresses, the $\Delta V_{\rm th}$ of both PMOS and NMOS devices increase, which in turn increases the path delay according to the relationship in Sec. \ref{sec: polynomial}. When the critical-path delay exceeds the clock period, $V_{\rm DD}$ is increased by $V_{\rm step}$ (set to 10 mV in this paper). This voltage increment reduces the path delay and restores the timing margin. The critical path delay then continues to increase over time until it again reaches the clock period, at which point the next $V_{\rm DD}$ adjustment is applied. This procedure is repeated throughout the 10-year lifetime. Fig.~\ref{fig:avs_fault_tolerance} demonstrates the curves of $V_{\rm DD}$, critical-path delay, and aging-induced degradation over time during AVS.

%% file: docs/4_Optimization.tex
\section{Fault-Tolerant Voltage Scaling}
\label{sec: optimization}

In the AVS process described above, increasing $V_{\rm DD}$ accelerates device aging, which then forces more frequent voltage boosts, forming a positive feedback loop. This loop can be mitigated by permitting controlled timing violations (i.e., increasing the maximum tolerable delay threshold (delay\_max) that triggers a voltage adjustment), thereby reducing unnecessary $V_{\rm DD}$ increases, slowing aging, and lowering lifetime energy consumption. In this section, we target AI accelerators and exploit the inherent error resilience of DNNs to curb voltage escalation. Specifically, we quantify (i) how delay\_max maps to the resulting BER under timing violations and (ii) how BER degrades model accuracy.


\subsection{Error Modeling}
Permitting timing violations mitigates $V_{\rm DD}$ increases and alleviates aging, albeit at the cost of induced computational errors. Consequently, it is essential to characterize the relationship between the delay\_max and the resulting BER. As discussed in Sec. \ref{sec: delta_Vth}, we adopt uniform aging model as a high-fidelity first-order approximation. Under this model, delay\_max serves as a global aging indicator, with all path delays scaling proportionally. We define a timing error as an aged path delay exceeding the nominal clock period, and we calculate BER by counting timing errors under a realistic workload described in Sec. \ref{sec: delta_Vth}.

\vspace{-6pt}
\subsection{Error Impact}
\vspace{-2pt}

Since BER can degrade model accuracy, quantifying its impact is essential for leveraging the inherent fault tolerance of neural networks. Prior works listed in Sec. \ref{sec: fault tolerance of DNN} have extensively characterized error resilience across different neural networks. Using these resilience characterizations, we derive the maximum tolerable BER for a target accuracy loss and map it to the corresponding delay\_max.
For example, \cite{xie2025realm} shows that O and Down projections are more sensitive to errors. Thus, we can assign tighter delay\_max to these operations.
This approach enables adaptive voltage-scaling strategies that are optimized based on model-level sensitivity to timing errors.


%% file: docs/6_Evaluation.tex
\vspace{-5pt}
\section{Evaluation}
\label{sec: evaluation}
\vspace{-5pt}
\subsection{Experiment Setup}

The hardware evaluation maps aging and voltage conditions to BERs using the timing error analysis framework introduced in Sec. \ref{sec:avs}. We implement RTL for 256×256 systolic arrays with 8-bit inputs and 32-bit outputs. All designs are synthesized using Synopsys Design Compiler with a commercial 14nm PDK (0.9V, 25°C) consistent with aging compact models. We report the hardware metrics, such as area and power, based on the synthesis results. The clock period is set to 1.6 ns, with a nominal critical path delay of 1.542 ns. Path delays are extracted using PrimeTime. We use the fitted model of the critical path in Sec. \ref{sec: polynomial} to estimate delay degradation proportion versus $\Delta V_{\rm th}$ and $V_{\rm DD}$ for all paths. 
We use the \llm~model \cite{grattafiori2024llama} evaluated on LAMBADA \cite{paperno2016lambada} as a case study to demonstrate our AVS framework, allowing up to a 0.5\% accuracy degradation. We emphasize that the proposed framework readily extends to other applications by parameterizing the acceptable timing-violation level.


\begin{table}[htbp]
\centering
\renewcommand{\arraystretch}{1.1}
\caption{Comparison of Aging Evaluation for AVS}
\begin{tabular}{c c c  r r r  r}
\toprule
\multirow{2}{*}{{$V_{\rm DD}$}} & \multirow{2}{*}{{Recovery}} & \multirow{2}{*}{{AVS}} & \multicolumn{3}{c}{{PMOS  (mV)}} & {NMOS} \\
 & & & {HCI} & {BTI} & {Total} & {(mV)} \\
\midrule
$V_{\rm nom}$       & \ding{55} & \ding{55} & 19.8 & 62.2  & 82.0  & 50.5 \\
$V_{\rm nom}$        & \ding{51} & \ding{55} & 18.2 & 54.9  & 73.1  & 46.1 \\ \hline
$V_{\max}$         & \ding{55} & \ding{55} & 27.3 & 103.4 & 130.7 & 105.2 \\
$V_{\rm nom}\rightarrow V_{\max}$    & \ding{51} & \ding{51} & 23.7 & 81.6  & 105.3 & 85.1 \\
\bottomrule
\end{tabular}
\label{tab:AVS}
\end{table}

\subsection{AVS Comparison with Previous Evaluation}
By accounting for recovery effects and historical voltage changes, our aging evaluation in the AVS process yields more accurate results. Table~\ref{tab:AVS} presents the estimated aging degradation across different approaches. Under nominal voltage ($V_{\rm nom}=0.90\rm V$), incorporating recovery mechanisms reduces $\Delta V_{\rm th}$ by 10.9\% for PMOS (from 82.0 mV to 73.1 mV) and 8.7\% for NMOS (from 50.5 mV to 46.1 mV), highlights the importance of accurate aging modeling. In a 10-year AVS scenario, $V_{DD}$ increases from 0.90 V to 1.02 V. Compared to prior works \cite{chan2014aging} that estimate aging based on constant maximum voltage ($V_{\rm max}=1.02\rm V$), our model considers voltage changes and reduces estimated $\Delta V_{\rm th}$ by 19.4\% for PMOS (from 130.7 mV to 105.3 mV) and 19.1\% for NMOS (from 105.2 mV to 85.1 mV), demonstrating a significant reduction in overestimation.

\begin{figure*}[!tb]
    \centering
    \includegraphics[width=0.9\linewidth]{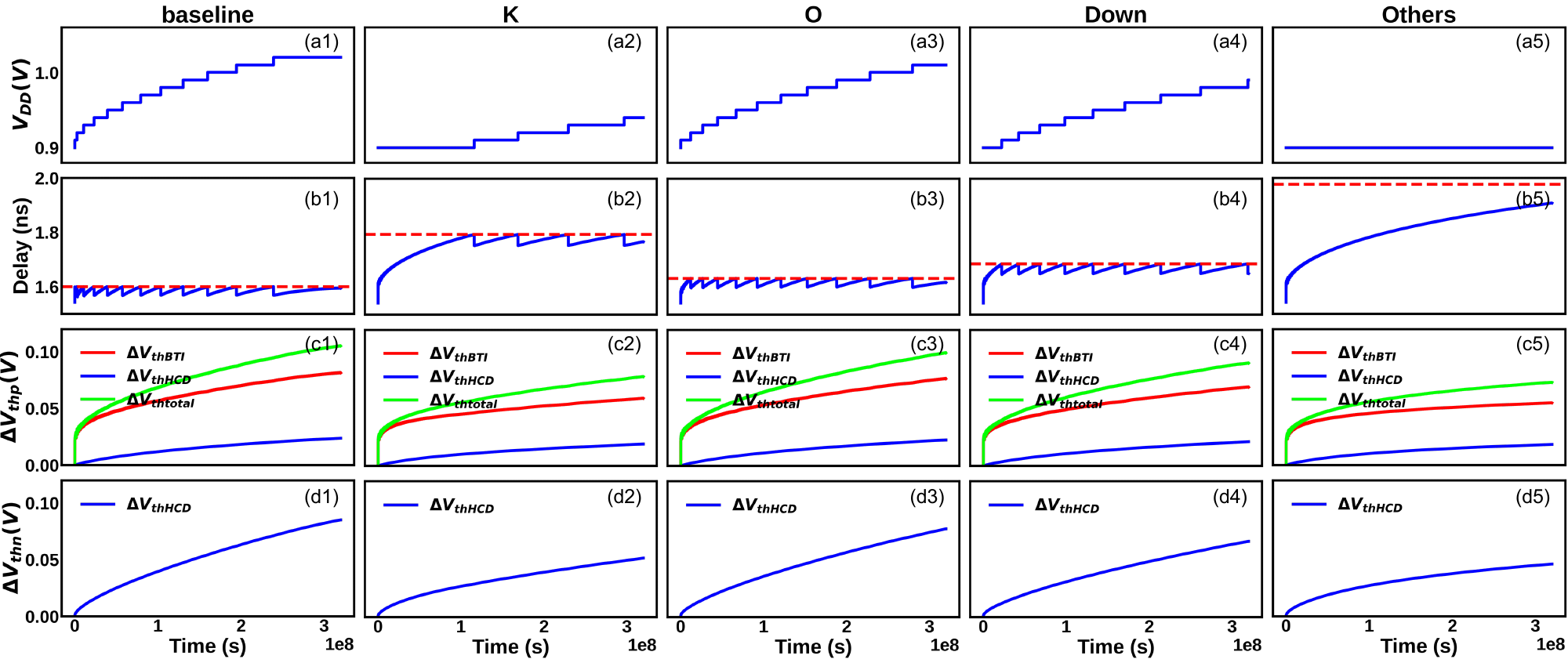}
    \vspace{-10pt}
    \caption{AVS process with and without fault tolerance, considering components \texttt{K}, \texttt{O}, \texttt{Down}, and others (\texttt{Q}, \texttt{V}, \texttt{QK}$^\texttt{T}$, \texttt{SV}, \texttt{Gate}, and \texttt{Up}). (a) $V_{\mathrm{DD}}$ over time. Allowing timing violations significantly reduces voltage escalation. (b) Critical path delay over time. Relaxed timing constraints lower voltage increase requirements. (c) $\Delta V_{\mathrm{th}}$ of PMOS over time. (d) $\Delta V_{\mathrm{th}}$ of NMOS over time.}
    \label{fig:avs_fault_tolerance}
    \vspace{-10pt}
\end{figure*}

\begin{table}[tbp]
\centering
\renewcommand{\arraystretch}{1.1}
\caption{$\Delta V_{\rm th}$ Reduction and Power Saving over Lifetime}
\begin{tabular}{ccccccc}
\toprule
Network&$V_{\rm final}$&$\Delta V_{\rm th, p}$&$\Delta V_{\rm th, n}$&$V_{\rm eff}$&$P_{\rm avg}$&Power    \\
Component&(V)&(mV)&(mV)&(V)&(W)&Saving    \\

\midrule
None    & 1.02 & 105.3 & 85.1 & 0.99 & 1.03  & /        \\
\midrule
\texttt{Q}       & 0.90 & 73.1  & 46.1 & 0.90 & 0.85 & 17.0\%  \\
\texttt{K}       & 0.94 & 79.0  & 52.1 & 0.92 & 0.88 & 14.3\%  \\
\texttt{V}       & 0.90 & 73.1  & 46.1 & 0.90 & 0.85 & 17.0\%  \\
\texttt{QK}$^{\texttt{T}}$      & 0.90 & 73.1  & 46.1 & 0.90 & 0.85 & 17.0\%  \\
\texttt{SV}      & 0.90 & 73.1  & 46.1 & 0.90 & 0.85 & 17.0\%  \\
\texttt{O}       & 1.01 & 99.7  & 77.8 & 0.97 & 1.00 & 3.1\%   \\
\texttt{Gate}    & 0.90 & 73.1  & 46.1 & 0.90 & 0.85 & 17.0\%  \\
\texttt{Up}     & 0.90 & 73.1  & 46.1 & 0.90 & 0.85 & 17.0\%  \\
\texttt{Down}    & 0.99 & 90.8  & 66.7 & 0.95 & 0.95 & 7.8\%   \\
\midrule
Avg. & /    & /     & /    & /     & 0.89 & \textbf{14.0\%}  \\
\bottomrule
\end{tabular}
\label{tab:lifetime}
\end{table}


\subsection{Fault Tolerance Impact on AVS}
This subsection demonstrates how tolerating timing violations, based on the error resilience of specific LLM components, affects AVS. We first conduct error injection experiments to calibrate the maximum tolerable BER that results in no more than 0.5\% performance degradation. We then derive the corresponding path delay increase causing this BER. Fig.~\ref{fig:avs_fault_tolerance}(a) presents the  $V_{\mathrm{DD}}$ curve over the lifetime for baseline (no fault tolerance) and components \texttt{K}, \texttt{O}, \texttt{Down}, and a group of other components. For the other components (\texttt{Q}, \texttt{V}, \texttt{QK}$^\texttt{T}$, \texttt{SV}, \texttt{Gate}, and \texttt{Up}), path delay never reaches the maximum tolerable threshold. Fig.~\ref{fig:avs_fault_tolerance}(b) further shows the corresponding critical path delays, with the maximum tolerable delay highlighted by a red dotted line. These results demonstrate that selectively permitting timing violations can substantially reduce the number of required voltage increases throughout the AVS lifetime.

\subsection{Lifetime Aging and Power Optimization}

Relaxing timing constraints slows $V_{DD}$ increases, which in turn alleviates aging degradation, as shown in Fig.~\ref{fig:avs_fault_tolerance}(c) for PMOS and (d) for NMOS. The curves also reveal that $V_{DD}$ increases accelerate aging. Table~\ref{tab:lifetime} summarizes the final threshold voltage shift across all network components. For PMOS,$\Delta V_{\rm th, p}$ is reduced by up to 30.6\%, from 105.3 mV to 73.1 mV, while NMOS degradation $\Delta V_{\rm th, n}$ is reduced by up to 45.8\%, from 85.1 mV to 46.1 mV. These reductions enhance device reliability and extend potential lifetime. We further estimate the effective voltage and average power consumption of our synthesized systolic array over the lifetime. Our approach achieves a 14.0\% total power savings on average over the entire lifetime.

%% file: docs/7_Conclusion.tex
\section{Conclusion}

This paper presents an accurate aging evaluation framework for AVS systems that incorporates advanced BTI/HCI models with historical voltage dependency and iterative 10-year lifetime extrapolation. 
To mitigate the positive feedback between aging and voltage increase, we propose a fault-tolerant voltage scaling policy that permits bounded timing violations and defers voltage adjustments by exploiting DNN error resilience. Compared with conventional maximum-voltage-based evaluation, our framework substantially reduces aging overestimation. Using \llm~as a case study, we show that the proposed policy preserves model accuracy while avoiding unnecessary voltage scaling, achieving up to 45.8\% reduction in NMOS $\Delta V_{\rm th}$, 30.6\% in PMOS, and 14.0\% average lifetime power savings.
